\documentclass[showpacs,10pt,twocolumn,prb]{revtex4-1}
\usepackage{amsmath}
\usepackage{amssymb}
\usepackage{graphics}
\usepackage{epsfig}

\begin{document}


\title{Critical fields, thermally-activated transport and critical current
density of $\beta $-FeSe single crystals}
\author{Hechang Lei, Rongwei Hu,$^{*}$ and C. Petrovic}
\affiliation{Condensed Matter Physics and Materials Science Department, Brookhaven
National Laboratory, Upton, NY 11973, USA}
\date{\today}

\begin{abstract}
We present critical fields, thermally-activated flux flow (TAFF) and
critical current density of tetragonal phase $\beta $-FeSe single crystals.
The upper critical fields $H_{c2}(T)$ for H$\parallel $(101) and H$\bot $%
(101) are nearly isotropic and are likely governed by Pauli limiting
process. The obtained large Ginzburg-Landau parameter $\kappa $ $\sim $
72.3(2) indicates that $\beta $-FeSe is a type-II superconductor with
smaller penetration depth than in Fe(Te,Se). The resistivity below $T_{c}$
follows Arrhenius TAFF behavior. For both field directions below 30 kOe
single vortex pinning is dominant whereas collective creep becomes important
above 30 kOe. The critical current density $J_{c}$ from M-H loops for H$%
\parallel $(101) is about five times larger than for H$\bot $(101), yet much
smaller than in other iron-based superconductors.
\end{abstract}

\maketitle

\section{Introduction}

Among iron-based superconductors, the tetragonal $\beta $-FeSe\ has trigged
great interest because of simplest structure and superconductivity below
about 8 K without any carrier doping.\cite{Hsu FC} Besides $\beta $-FeSe,
the superconductivity has also been discovered in binary Fe(Te,Se) and
Fe(Te,S) materials.\cite{Yeh KW}$^{,}$\cite{Mizuguchi} These materials also
have similar Fermi surface to FePn (Pn = P, As)-based superconductors,\cite%
{Subedi} even though they have FeCh (Ch = Te, Se, S) layers stacked along
the c-axis as opposed to FePn layers. Understanding this similarity is
rather important. Binary FeCh material, $\beta $-FeSe has another notable
characteristics. Application of pressure leads to a significant enhancement
of $T_{c}$ up to 37 K at around 9 GPa, the third highest known critical
temperature for any binary compound.\cite{Medvedev}

In order to study anisotropic and intrinsic physical properties of
materials, single crystals are required. When compared to Fe(Te,Se) and
Fe(Te,S), extremely complex phase diagram of FeSe makes impurities such as $%
\alpha $-FeSe and Fe$_{7}$Se$_{8}$ ubiquitous in as-grown crystals, or
sometimes polycrystals impeding the understanding of $\beta $-FeSe.\cite%
{Zhang SB}$^{-} $\cite{McQueen}

Here, we report intrinsic superconducting properties of $\beta $-FeSe single
crystals. These include critical fields $H_{c2}$ and $H_{c1}$,
thermally-activated flux flow (TAFF) behavior, and the critical current
density $J_{c}$. Our results show that $\beta $-FeSe is a type-II
superconductor with large Ginzburg-Landau parameter $\kappa $ and smaller
critical current density when compared to other iron-based superconductors.
Single vortex pinning dominates vortex dynamics below 30 kOe, whereas
collective creep becomes important at higher magnetic fields.

\section{Experiment}

Details of synthesis and structural characterization are explained elsewhere.%
\cite{Hu RW2} Thin Pt wires were attached to electrical contacts made of
Epotek H20E silver epoxy for a standard four-probe measurement, with current
flowing in crystal plane. Sample geometry were measured with an optical
microscope Nikon SMZ-800 with 10 $\mu $m resolution. Magnetization and
resistivity measurements were carried out in Quantum Design MPMS and PPMS,
respectively.

\section{Results and Discussion}

\begin{figure}[tbp]
\centerline{\includegraphics[scale=0.7]{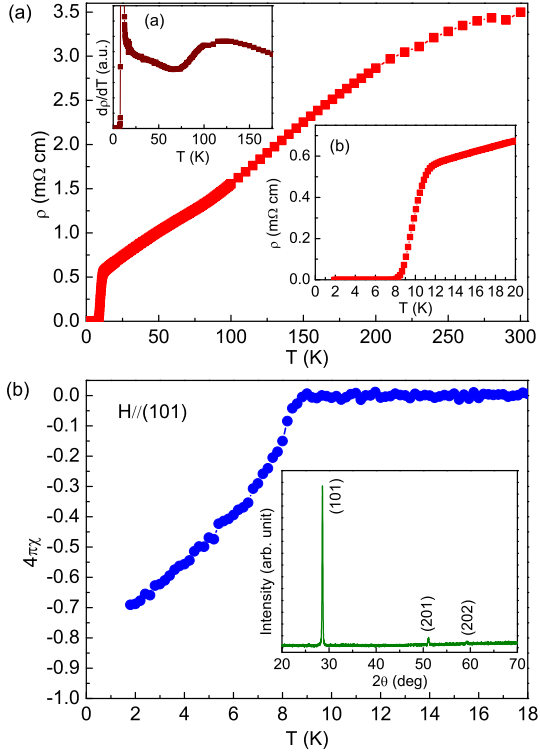}} \vspace*{-0.3cm}
\caption{(a) Temperature dependence of the in-plane resistivity $\protect%
\rho (T)$ of $\protect\beta $-FeSe single crystals. Inset (a) shows
derivative of resistivity data $d\protect\rho /dT$ as a function of
temperature. Inset (b) shows enlarged resistivity curve near $T_{c}$. (b)
Temperature dependence of dc magnetic susceptibility of $\protect\beta $%
-FeSe single crystals for H = 10 Oe along the (101) plane. Inset: single
crystal XRD pattern of $\protect\beta $-FeSe.}
\end{figure}

Figure 1(a) shows the temperature dependence of the in-plane resistivity $%
\rho (T)$ below 300 K. The residual resistivity ratio (RRR) of 14 is double
that of the hexagonal shape crystals,\cite{Zhang SB} indicating good sample
quality. The curvature of $\rho (T)$ changes at about 100 K (Fig. 1(a) inset
(a)) due to structural and magnetic transitions in agreement with previous
results.\cite{Hsu FC}$^{,}$\cite{Zhang SB}$^{,}$\cite{Margadonna} With
further decrease in temperature, superconductivity emerges with $T_{c,onset}$
$\simeq $ 11.4 K and $T_{c,0}$ $\simeq $ 8.1 K (Fig. 1(a) inset (b)),
similarly to reported values in the literature.\cite{Zhang SB} Fig. 1(b)
presents the ac susceptibility of FeSe single crystal for field
perpendicular to the crystal plane. The corresponding superconducting volume
fraction at T = 1.8 K is about 70 \%, confirming the bulk nature of
superconductivity. The XRD pattern of a single crystal (Fig. 1(b) inset)
reveals that the crystal surface is normal to (101) direction, also similar
to previous results.\cite{Zhang SB}$^{,}$\cite{Mok BH}

\begin{figure}[tbp]
\centerline{\includegraphics[scale=0.8]{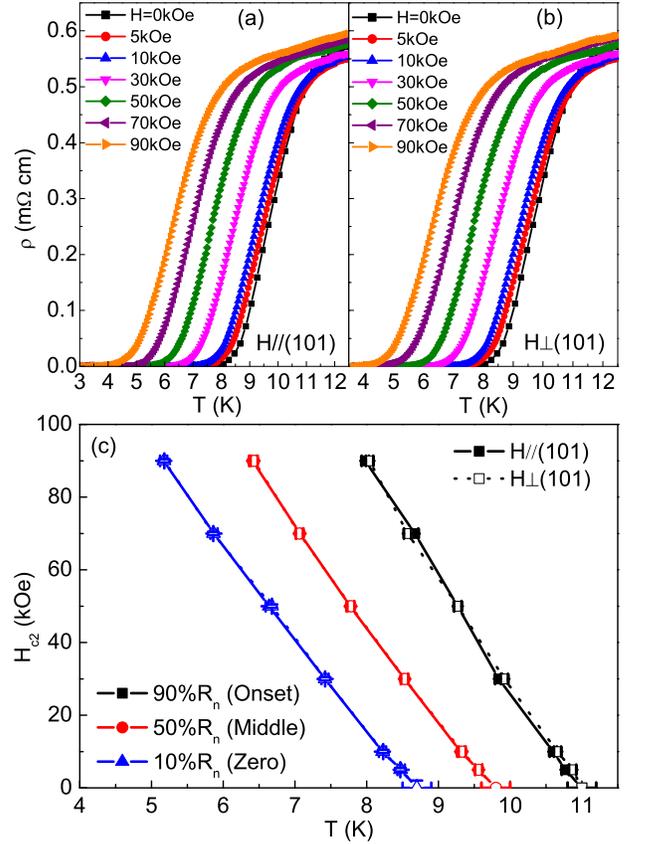}} \vspace*{-0.3cm}
\caption{Temperature dependence of the resistivity $\protect\rho (T)$ of $%
\protect\beta $-FeSe single crystals for (a) H(101) and (b) H(101) in the
magnetic field up to 90 kOe. (c) Temperature dependence of the resistive
upper critical field $H_{c2}(T)$ corresponding\ to three defined
temperatures for both field directions (see text).}
\end{figure}

Figure 2 (a) and (b) show temperature dependence of $\rho (T)$ in various
fields for H$\Vert $(101) and H$\bot $(101). The $T_{c}$ shifts to lower
temperature without obvious broadening for both directions with the increase
in magnetic fields. Temperature dependence of the upper critical fields $%
H_{c2}(T)$ was determined from the resistivity drops to 90\%, 50\%, and 10\%
of the normal-state resistivity $\rho _{n}$(T,H) (Fig. 2(c)). The
normal-state resistivity was determined by linearly extrapolating the
normal-state behavior above the onset of superconductivity. The $H_{c2}(T)$
curves are nearly linear and the initial slopes $dH_{c2}/dT|_{T_{c}}$ are
given in Table 1. The slopes are nearly identical for both field directions
and the anisotropy of the upper critical field $\gamma (T)=H_{c2,H\bot
(101)}(T)/H_{c2,H\Vert (101)}(T)$ is also isotropic within the experimental
error.

\begin{table*}[tbp] \centering%
\caption{Superconducting parameters of $\beta$-FeSe.}%
\begin{tabular}{cccccccccccc}
\hline\hline
& $T_{c,mid}$ & \multicolumn{3}{c}{$(dH_{c2}/dT)_{T_{c}}$} & $H_{c2,mid}(0)$
& $\xi (0)$ & $H_{P}(0)$ & $H_{c1}(0)$ & $\lambda (0)$ & $H_{c}(0)$ & $%
\kappa (0)$ \\
& (K) & \multicolumn{3}{c}{(kOe/K)} & (kOe) & (nm) & (kOe) & (Oe) & (nm) &
(kOe) &  \\
&  & Onset & Middle & Zero &  &  &  &  &  &  &  \\ \hline
H$\parallel $(101) & 9.8(2) & -30.3(6) & -26.5(6) & -25.4(4) & 180(4) &
4.28(5) & 180(4) & 75(2) & 309(4) & 1.76(3) & 72.3(2) \\
H$\perp $(101) & 9.8(2) & -29.6(6) & -26.6(6) & -25.5(4) & 180(4) & 4.28(5)
& 180(4) &  &  &  &  \\ \hline\hline
\end{tabular}%
\label{TableKey copy(1)}%
\end{table*}%

Within the weak coupling BCS theory\cite{Werthamer} and using the slope
determined from the midpoint of resistive transition with $T_{c}$ = 9.8 K we
can estimate $H_{c2}$(0) = -0.693$T_{c}$($dH_{c2}/dT|_{T_{c}}$) = 180(4) kOe
for both field directions. The results are close to the Pauli paramagnetic
limit $H_{P}$(0) = 1.84$T_{c}$ = 180(4) kOe.\cite{Clogston} It implies that
the spin-paramagnetic effect may be the dominant pair-breaking mechanism in
FeSe for both field directions, similar to Fe(Te,Se) and Fe(Te,S).\cite{Lei
HC1}$^{,}$\cite{Lei HC2} The superconducting coherence length $\xi $(0)
estimated using the Ginzburg-Landau formula $H_{c2}(0)=\Phi _{0}/2\pi \xi
^{2}(0)$, where $\Phi _{0}$ = 2.07$\times $10$^{-15}$ Wb is the flux
quantum, is $\xi $(0) = 4.28(5) nm, which is somewhat larger than Fe(Te,Se)
and Fe(Te,S).\cite{Lei HC1}$^{,}$\cite{Lei HC2}

\begin{figure}[tbp]
\centerline{\includegraphics[scale=0.45]{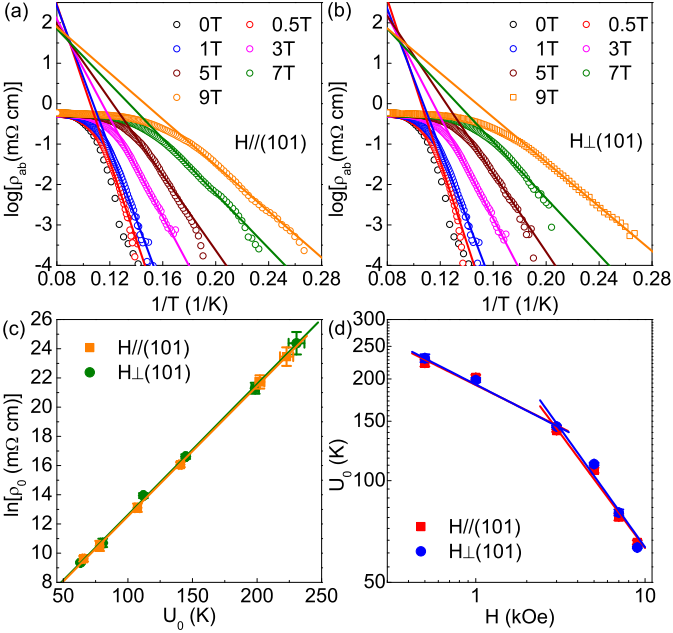}} \vspace*{-0.3cm}
\caption{Log$\protect\rho (T,H)$ vs. $1/T$ in various field for (a) H$\Vert $%
(101) and (b) H$\bot $(101). The corresponding solid lines are fitting
results from the Arrhenius relation. (c) $ln\protect\rho _{0}(H)$ vs. $%
U_{0}(H)$ derived from Arrhenius relation for both field directions. The
solid lines are linear fitting results. (d) Field dependence of $U_{0}(H)$.
The solid lines are power-law fitting using $U_{0}(H)\sim H^{-\protect\alpha %
}$.}
\end{figure}

According to the thermally-activated flux flow (TAFF) theory, the $ln\rho
-1/T$ in TAFF region can be described using Arrhenius relation,\cite%
{Palstra1}$^{,}$\cite{Palstra2}

\begin{equation}
ln\rho (T,H)=ln\rho _{0}(H)-U_{0}(H)/T
\end{equation}

where $ln\rho _{0}(H)=ln\rho _{0f}+U_{0}(H)/T_{c}$ is the
temperature-independent constant and $U_{0}(H)$ is the apparent activated
energy. Hence, the $ln\rho (T,H)$ vs. $1/T$ should be linear in TAFF region.
As shown in Fig. 3 (a) and (b), the Arrhenius relation (solid lines) can fit
the experimental data very well for both field directions. The results are
shown in the common logarithmic scale in the figures, but we calculate them
in the natural one. The obtained $U_{0}$ are similar for both field
directions. They are comparable to that in Fe(Te,S) and much smaller than in
Fe(Te,Se).\cite{Lei HC3}$^{,}$\cite{Yadav} The good linear behavior
indicates that the temperature dependence of thermally activated energy
(TAE) $U(T,H)$ is approximately linear, i.e., $U(T,H)=U_{0}(H)(1-T/T_{c})$.%
\cite{Palstra1}$^{,}$\cite{Palstra2} The $\log \rho (T,H)$ lines for
different fields extrapolate to the same temperature $T_{cross}$, which
should equal to $T_{c}$.\cite{Lei HC3} The extrapolated temperatures are
about 11.1 K for both H$\Vert $(101) and H$\bot $(101). Moreover, $ln\rho
_{0}(H)-U_{0}(H)$ show linear behavior for both field directions (Fig.
3(c)). Fits using $ln\rho _{0}(H)=ln\rho _{0f}+U_{0}(H)/T_{c}$, yielded
values of $\rho _{0f}$ and $T_{c}$ 37(1) $m\Omega \cdot cm$ and 11.2(1) K
for H$\Vert $(101) and 39(2) $m\Omega \cdot cm$ and 11.2(1) K for H$\bot $%
(101). The $T_{c}$ values are consistent with the values of $T_{cross}$
within the error bars. The $U_{0}(H)$ shows a power law ($U_{0}(H)\sim
H^{-\alpha }$) field dependence for both directions (Fig. 3(d)). For H$\Vert
$(101), $\alpha $ = 0.25(6) for H $<$ 30 kOe and $\alpha $ = 0.68(6) for H $%
> $ 30 kOe; For H$\bot $(101), $\alpha $ = 0.26(2) for H $<$ 30 kOe and $%
\alpha $ = 0.70(9) for H $>$ 30 kOe. The weak power law decreases of $%
U_{0}(H)$ in low fields for both field directions implies that single-vortex
pinning dominates in this region,\cite{Blatter} followed by a quicker
decrease of $U_{0}(H)$ in field which could be related to a crossover to a
collective flux creep regime.\cite{Yeshurun}

\begin{figure}[tbp]
\centerline{\includegraphics[scale=0.8]{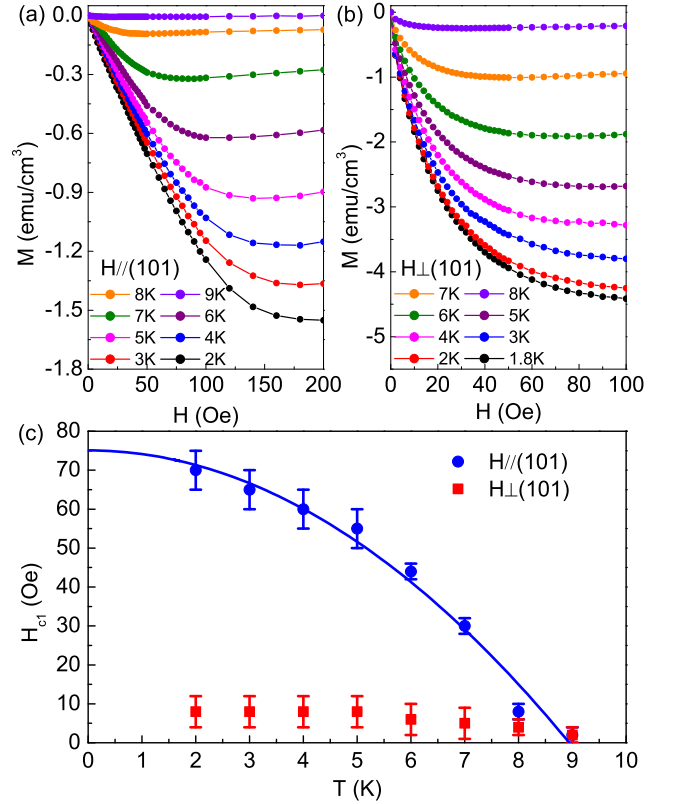}} \vspace*{-0.3cm}
\caption{Low field M(H) of $\protect\beta $-FeSe single crystals at various
temperatures for (a) H$\parallel $(101) and H$\bot $(101). (c) Temperature
dependence of $H_{c1}(T)$ for both field directions. The solid line is the
fitted lines using $H_{c1}(T)=H_{c1}(0)[1-(T/T_{c}){^{2}}]$ for H$\parallel $%
(101).}
\end{figure}

Low field M(H) at various temperatures for H$\parallel $(101) and H$\bot $%
(101) are shown in Fig. 4(a) and (b), respectively. All curves exhibit
linear behavior for low fields and then deviate from linearity at different
field values for different temperatures. The values of $H_{c1}$ are
determined by examining the point of deviation from the linear slope of the
magnetization curve. The temperature dependence of $H_{c1}(T)$ for both
field directions are shown in Fig. 4(c). For H$\parallel $(101), the $%
H_{c1}(T)$ can be well fitted using the formula $%
H_{c1}(T)=H_{c1}(0)[1-(T/T_{c})^{2}]$ and the obtained $H_{c1,H\parallel
(101)}$(0) is 75(2) Oe.\ For H$\bot $(101), it is difficult to estimate the $%
H_{c1,H\bot (101)}(T)$ and obtain reliable fits. This is due to the small
obtained value for $H_{c1}$ with relative large error and significant
demagnetization factors.

Since Fe(Te,Se) and Fe(Te,S) superconductors are in the dirty limit, we
assume the same for FeSe.\cite{Kim} Using estimated values of $%
H_{c1,H\parallel (101)}$(0), $H_{c2}$(0) and $\xi $(0), we evaluate
additional parameters using expressions $H_{c2}$(0) = $\sqrt{2}\kappa H_{c}$%
(0) and $H_{c1}$(0) = $\frac{H_{c}(0)}{\sqrt{2}\kappa }(\ln \kappa +0.08)$
where $\kappa $ = $\lambda /\xi $ is the Ginzburg-Landau (GL) parameter and $%
H_{c}$(0) is the thermodynamic upper critical field at T = 0 K.\cite%
{Abrikosov} We obtain $\kappa _{H\Vert (101)}^{\text{ }}$(0) = 72.3(2), $%
H_{c,H\parallel (101)}$(0) = 1.76(3) kOe, and penetration depth for H$%
\parallel $(101) $\lambda _{H\parallel (101)}$(0) = 309(4) nm which is
somewhat smaller than in Fe(Te,Se).\cite{Kim} All superconducting parameters
are listed in Table 1.

\begin{figure}[tbp]
\centerline{\includegraphics[scale=0.45]{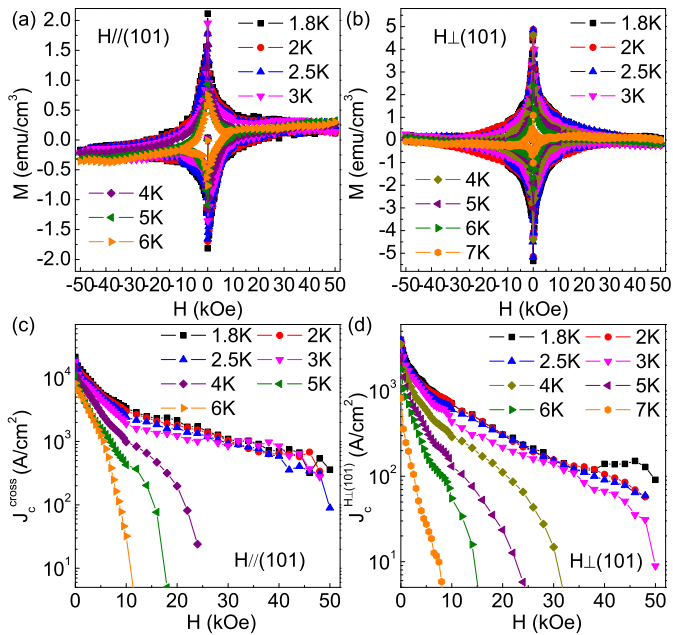}} \vspace*{-0.3cm}
\caption{M(H) loops of $\protect\beta $-FeSe single crystals at various
temperatures with field up to 50 kOe for (a) H$\parallel $(101) and H$\bot $%
(101). (c) In-plane and (d) interplane superconducting critical currents as
determined from magnetization measurements using the Bean model.}
\end{figure}

Figures 5(a) and (b) show isothermal magnetization curves M(H) at various
temperatures for H$\parallel $(101) and H$\bot $(101). The shape of the M(H)
curves confirms that $\beta $-FeSe is a typical type-II superconductor. The
data exhibit a central peak at zero magnetic field and then the
magnetization decreases continuously with increasing magnetic field. On the
other hand, there is a weak ferromagnetic (WFM) background superimposed on
the superconducting M(H) curve for H$\parallel $(101). This WFM background
possibly arises due to vacancy-induced magnetic cluster.\cite{Hu RW2} From
the irreversible parts of the M(H) loops, the critical current can be
determined using the Bean model.\cite{Bean}$^{,}$\cite{Gyorgy} For a
rectangular-shaped crystal with dimension c $<$ a $<$ b, when H$\bot $(101),
the in-plane critical current density $J_{c}^{H\bot (101)}(H)$ is given by $%
J_{c}^{H\bot (101)}(H)=20\Delta M(H)/[a(1-a/3b)]$, where a and b (a $<$ b)
are the in-plane sample sizes, $\Delta M(H)$ is the width of the magnetic
hysteresis loop in emu/cm$^{3}$. On the other hand, for H$\parallel $(101),
there are two different current densities: the vortex motion across the
planes, $J_{c}^{cross}(H)$ and that parallel to the planes, $J_{c}^{para}(H)$%
. Assuming $a,b\gg (c/3)\cdot J_{c}^{para}(H)/J_{c}^{cross}(H)$,\cite{Gyorgy}
we obtain $J_{c}^{cross}(H)\approx 20\Delta M(H)/c$. The magnetic field
dependence of $J_{c}^{cross}(H)$ and $J_{c}^{H\bot (101)}(H)$ is shown in
Fig. 4(c) and 4(d), respectively. It should be noted that for H$\parallel $%
(101), the WFM background has only minor effect on the calculation of $\Delta
M(H)$, because of very weak moment contribution of WFM. The error of
calculated $J_{c}^{cross}(H)$ is about 10\% without subtracting WFM
background. It can be seen that the $J_{c}^{cross}(0)$ and $J_{c}^{H\bot
(101)}(0)$\ at 1.8 K are about 2.2$\times $10$^{4}$ and 4$\times $10$^{3}$
A/cm$^{2}$, which are much smaller than in Fe(Te,Se) and Fe(Te,S) at the
same temperature.\cite{Taen}$^{,}$\cite{Hu RW} The ratio of $%
J_{c}^{cross}(H)/J_{c}^{H\bot (101)}(H)$ is about 5.4 at 1.8 K. Moreover,
above 4 K, the critical current densities decrease with the applied field
more quickly than below 4 K, suggesting that the pinning mechanism may
change at 4 K.

\section{Conclusion}

In summary, we present the superconducting properties of $\beta $-FeSe
single crystals. The results indicate that the $H_{c2}(T)$ is isotropic and
the spin-paramagnetic effect may be the dominant pair-breaking mechanism for
H$\parallel $(101) and H$\bot $(101). The calculated GL parameter $\kappa $
indicates that $\beta $-FeSe is a typical type-II superconductor with the
large $\kappa $. The resistivity exhibits clear Arrhenius TAFF behavior with
a crossover from single vortex pinning region to collective creep region for
both field directions. On the other hand, the critical current density for
field along (101) plane is about five times larger than for field normal to
(101) plane but still smaller than in Fe(Te,Se) and Fe(Te,S) at the same
temperature.

\section{Acknowledgements}

Work at Brookhaven is supported by the U.S. DOE under Contract No.
DE-AC02-98CH10886 and in part by the Center for Emergent Superconductivity,
an Energy Frontier Research Center funded by the U.S. DOE, Office for Basic
Energy Science.

* Present address: Ames Laboratory US DOE and Department of Physics and
Astronomy, Iowa State University, Ames, IA 50011, USA


\begin{thebibliography}{99}
\bibitem{Hsu FC} F. C. Hsu, J. Y. Luo, K. W. Yeh, T. K. Chen, T. W. Huang,
P. M. Wu, Y. C. Lee, Y. L. Huang, Y. Y. Chu, D. C. Yan, and M. K. Wu, Proc.
Natl. Acad. Sci. USA \textbf{105}, 14262 (2008).

\bibitem{Yeh KW} K.-W. Yeh, T. W. Huang, Y. L. Huang, T. K. Chen, F. C. Hsu,
P. M. Wu, Y. C. Lee, Y. Y. Chu, C. L. Chen, J. Y. Luo, D. C. Yan, and M. K.
Wu, EPL \textbf{84}, 37002 (2008).

\bibitem{Mizuguchi} Y. Mizuguchi, F. Tomioka, S. Tsuda, T. Yamaguchi, and Y.
Takano, Appl. Phys. Lett. \textbf{94}, 012503 (2009).

\bibitem{Subedi} A. Subedi, L. Zhang, D. J. Singh, and M. H. Du, Phys. Rev.
B \textbf{78}, 134514 (2008).

\bibitem{Medvedev} S. Medvedev, T. M. McQueen, I. Trojan, T. Palasyuk, M. I.
Eremets, R. J. Cava, S. Naghavi, F. Casper, V. Ksenofontov, G. Wortmann, and
C. Felser, Nature Mater. \textbf{8,} 630 (2009).

\bibitem{Zhang SB} S. B. Zhang, X. D. Zhu, H. C. Lei, G. Li, B. S. Wang, L.
J. Li, X. B. Zhu, Z. R. Yang, W. H. Song, J. M. Dai, and Y. P. Sun,
Supercond. Sci. Technol. \textbf{22}, 075016 (2009).

\bibitem{Patel} U. Patel, J. Hua, S. H. Yu, S. Avci, Z. L. Xiao, H. Claus,
J. Schlueter, V. V. Vlasko-Vlasov, U. Welp, and W. K. Kwok, Appl. Phys.
Lett. \textbf{94}, 082508 (2009).

\bibitem{Mok BH} B. H. Mok, S. M. Rao, M. C. Ling, K. J. Wang, C. T. Ke, P.
M. Wu, C. L. Chen, F. C. Hsu, T. W. Huang, J. Y. Luo, D. C. Yan, K. W. Ye,
T. B. Wu, A. M. Chang, and M. K. Wu, Cryst. Growth Des. \textbf{9}, 3260
(2009).

\bibitem{Pomjakushina} E. Pomjakushina, K. Conder, V. Pomjakushin, M.
Bendele, and R. Khasanov, Phys. Rev. B \textbf{80}, 024517 (2009).

\bibitem{McQueen} T. M. McQueen, Q. Huang, V. Ksenofontov, C. Felser, Q. Xu,
H. Zandbergen, Y. S. Hor, J. Allred, A. J. Williams, D. Qu, J. Checkelsky,
N. P. Ong, and R. J. Cava, Phys. Rev. B \textbf{79}, 014522 (2009).

\bibitem{Hu RW2} R. W. Hu, H. C. Lei, M. Abeykoon, E. S. Bozin, S. J. L. Billinge,
J. B. Warren, T. Siegrist, and C. Petrovic, Phys. Rev. B \textbf{83}, 224502
(2011).

\bibitem{Margadonna} S. Margadonna, Y. Takabayashi, M. T. McDonald, K.
Kasperkiewicz, Y. Mizuguchi, Y. Takano, A. N. Fitch, E. Suard, and K.
Prassides, Chem. Comm. \textbf{43}, 5607 (2008).

\bibitem{Werthamer} N. R. Werthamer, E. Helfand, and P. C. Hohenberg, Phys.
Rev. \textbf{147}, 295 (1966).

\bibitem{Clogston} A. M. Clogston, Phys. Rev. Lett. \textbf{9}, 266 (1962).

\bibitem{Lei HC1} H.\ C. Lei, R. W. Hu, E. S. Choi, J. B. Warren, and C.
Petrovic, Phys. Rev. B\ \textbf{81}, 094518 (2010).

\bibitem{Lei HC2} H.\ C. Lei, R. W. Hu, E. S. Choi, J. B. Warren, and C.
Petrovic, Phys. Rev. B\ \textbf{81}, 184522 (2010).

\bibitem{Palstra1} T. T. M. Palstra, B. Batlogg, L. F. Schneemeyer, and J.
V. Waszczak, Phys. Rev. Lett. \textbf{61}, 1662 (1988).

\bibitem{Palstra2} T. T. M. Palstra, B. Batlogg, R. B. van Dover, I. F.
Schneemeyer, and J. V. Waszczak, Phys. Rev. B \textbf{41}, 6621 (1990).

\bibitem{Lei HC3} H. C. Lei, R.\ W. Hu, E. S. Choi, and C. Petrovic, Phys.
Rev. B \textbf{82}, 134525 (2010).

\bibitem{Yadav} C. S. Yadav and P. L. Paulose, New J. Phys. \textbf{11},
103046 (2009).

\bibitem{Blatter} G. Blatter, M. V. Feigel'man, V. B. Geshkenbein, A. I.
Larkin, and V. M. Vinokur, Rev. Mod. Phys. \textbf{66}, 1125 (1994).

\bibitem{Yeshurun} Y. Yeshurun and A. P. Malozemoff, Phys. Rev. Lett.
\textbf{60}, 2202 (1988).

\bibitem{Kim} H. Kim, C. Martin, R. T. Gordon, M. A. Tanatar, J. Hu, B.
Qian, Z. Q. Mao, R. W. Hu, C. Petrovic, N. Salovich, R. Giannetta, and R.
Prozorov, Phys. Rev. B \textbf{81}, 180503 (2010).

\bibitem{Abrikosov} A. A. Abrikosov, Zh. Eksp. Teor. Fiz. \textbf{32}, 1442 (1957) [Sov. Phys. JETP \textbf{5}, 1174 (1957)].

\bibitem{Bean} C. P. Bean, Phys. Rev. Lett. \textbf{8}, 250 (1962).

\bibitem{Gyorgy} E. M. Gyorgy, R. B. van Dover, K. A. Jackson, L. F.
Schneemeyer, and J. V. Waszczak, Appl. Phys. Lett. \textbf{55}, 283 (1989).

\bibitem{Taen} T. Taen, Y. Tsuchiya, Y. Nakajima, and T. Tamegai, Phys. Rev.
B \textbf{80}, 092502 (2009).

\bibitem{Hu RW} R. W. Hu, E. S. Bozin, J. B. Warren, and C. Petrovic, Phys.
Rev. B \textbf{80}, 214514 (2009).
\end{thebibliography}
\end{document}